\documentclass[preprint2,letterpaper]{aastex62}
\usepackage{graphicx}
\usepackage{amssymb}
\usepackage{amsmath}
\usepackage{epstopdf}
\usepackage{color}

\def \arcsec {\ensuremath{^{\prime\prime}}}
\def \psr {PSR\,J1119$-$6127}
\newcommand{\src}{PSR\,J1119$-$6127}
\newcommand{\source}{\src{}}
\def \nustar {\emph{NuSTAR}}
\def \swift {\emph{Swift}}
\def \xmm {\emph{XMM-Newton}}

\begin{document}

\title{THE 2016 OUTBURST OF \src{}:\linebreak
COOLING \& A SPIN-DOWN DOMINATED GLITCH}

\author{R.~F.~Archibald}
\affiliation{Department of Astronomy and Astrophysics, University of Toronto
	50 St. George Street, Toronto, ON M5S 3H4, Canada}
\correspondingauthor{R.~F.~Archibald}
\email{archibald@astro.utoronto.ca}

\author{V.~M.~Kaspi}
\affiliation{Department of Physics \& McGill Space Institute, McGill University, 3600 University Street, Montreal QC, H3A 2T8, Canada}

\author{S.~P.~Tendulkar}
\affiliation{Department of Physics \& McGill Space Institute, McGill University, 3600 University Street, Montreal QC, H3A 2T8, Canada}

\author{P.~Scholz}
\affiliation{National Research Council of Canada, Herzberg Astronomy and Astrophysics, Dominion Radio Astrophysical Observatory, P.O. Box 248, Penticton, BC V2A 6J9, Canada}

\keywords{stars: individual (PSR\,J1119$-$6127) -- stars: neutron}

\begin{abstract}
We report on the aftermath of a magnetar outburst from the young, high-magnetic-field radio pulsar PSR~J1119$-$6127 that occurred on 2016 July 27. We present the results of a monitoring campaign using the {\it Neil Gehrels Swift} X-ray Telescope, {\it NuSTAR}, and  {\it XMM-Newton}.
After reaching a peak luminosity of $\sim$300 times the quiescent luminosity, the pulsar's X-ray flux declined by factor of $\sim$50 on a time scale of several months.
The soft X-ray spectra are well described by a blackbody and a hard power-law tail. After an initial rapid decline during the first day of the outburst, we observe the blackbody temperature rising from $kT = 0.9$\,keV to 1.05\,keV during the first two weeks of the outburst, before cooling to 0.9\,keV. During this time,  the blackbody radius decreases monotonically by a factor of $\sim4$ over a span of nearly 200 days. 
We also report a heretofore unseen highly pulsed hard X-ray emission component, which fades on a similar timescale to the soft X-ray flux, as predicted by models of relaxation of magnetospheric current twists.
The previously reported spin-up glitch which accompanied this outburst was followed by a period of enhanced and erratic torque, leading to a net spin-down of $\sim3.5\times10^{-4}$\,Hz, a factor of $\sim$24 over-recovery.
We suggest that this and other radiatively loud magnetar-type glitch recoveries are dominated by magnetospheric processes, in contrast to conventional radio pulsar glitch recoveries which are dominated by internal physics.

\end{abstract}

\section{Introduction}

\src{} is a young ($\tau_c \equiv P/2\dot{P} < 2$\,kyr, where $P$ is the spin period) pulsar with a spin-inferred dipolar magnetic field strength of $B \equiv 3.2 \times 10^{19}\,\sqrt{P\dot{P}}$~G~$= 4\times10^{13}$\,G -- among the highest for known radio pulsars. While most of this pulsar's observed properties are consistent with those of prototypical rotation-powered pulsars, \src{}, in quiescence, has a high, $\sim0.2$\,keV, surface temperature \citep[e.g.][]{2005ApJ...630..489G, 2008ApJ...684..532S, 2012ApJ...761...65N}. 
This is similar to the high surface
temperatures measured for other
high-magnetic-field pulsars
\citep{2005ApJ...618L..41K, 2011ApJ...734...44Z, 2013ApJ...764....1O}
The abnormally high surface temperature in a high-magnetic-field radio pulsar led to predictions that such sources could exhibit magnetar-like behavior \citep{2005ApJ...618L..41K}.

On 2016 July 27, \src{} emitted several magnetar-like bursts that were detected by the {\it Neil Gehrels  Swift} Burst Alert Telescope (BAT), and the  {\it Fermi} Gamma-ray Burst Monitor (GBM) \citep{2016ApJ...829L..25G, 2016ApJ...829L..21A}. In the few days following these bursts, follow-up at X-ray energies with {\it Neil Gehrels Swift} X-ray Telescope (XRT) and {\it NuSTAR} showed that the unabsorbed 0.5--10-keV X-ray flux
of \src{} had increased by a factor of $\sim$300.
Moreover, a hard X-ray component suddenly appeared, with emission extending at least
up to $\sim$70~keV, a spectral behavior previously well established in many magnetars
\citep[see][for a review]{2017ApJS..231....8E}.
The source also underwent a contemporaneous spin-up glitch \citep{2016ApJ...829L..21A}. 
Thus, \src{} displayed a classic magnetar-like outburst, as previously predicted, despite its normal appearance as a radio pulsar in the two decades since its discovery \citep[e.g.][]{2000ApJ...541..367C, 2015MNRAS.447.3924A}.
Interestingly, its associated
pulsar wind nebula \citep{2003ApJ...591L.143G, 2012ApJ...754...96K}
showed evidence for morphological changes
post-outburst \citep{2017ApJ...850L..18B}.
The radio emission also was affected by the magnetar activity, initially becoming undetectable as a radio pulsar, before returning with a steeper radio spectrum, and a changed, multi-component pulse shape \citep{2017ApJ...834L...2M}.

Relaxations from magnetar outbursts have
been studied extensively \cite[see][for a recent review]{2018MNRAS.474..961C} and can be used to constrain models of magnetar physics.  Flux and spectral evolution models can constrain and/or test models of crustal cooling \citep[e.g.][]{2002ApJ...580L..69L} or of magnetospheric twisting \citep[e.g.][]{2009ApJ...703.1044B}, while timing evolution post-glitch in
radio pulsars can in principle constrain the structure and content of the neutron-star interior 
\citep[e.g.][]{1992Natur.359..616L}.  The occurrence of all these phenomena in one source may provide clues regarding interactions between neutron star interiors and exteriors; with all in a high-magnetic-field radio pulsar such as \src{}, we have the opportunity, in comparing with analogous behavior in {\it bona fide} magnetars, to see how such interactions depend on  field strength.

Here we report on the post-outburst evolution of the timing and spectral properties of \src{} as observed using the {\it Swift} XRT, {\it NuSTAR}, and  {\it XMM-Newton}. 
We show that the hard X-ray component, which we find to be highly pulsed, relaxed on approximately the same
time scale as did the soft X-ray emission, suggesting, for the first time observationally, a related physical origin for these two distinct spectral components.
We further show that the original spin-up glitch reported by \cite{2016ApJ...829L..21A} was followed by a period of increased $\dot{\nu}$, by up to a factor of 5 more than the normal value, leading to a net spin-down of $\sim3.5\times10^{-4}$\,Hz -- a value comparably large to that seen following the giant flare in SGR\,1900$+$14 \citep{2000ThompsonSpinDown}. 

\section{Observations and Data Analysis}
\label{sec:obs}

\subsection{Swift Observations}
\label{sec:xrt}
We downloaded Level 1 XRT data data products from the HEASARC \emph{Swift} archive, and reduced them using the  {\tt xrtpipeline} standard reduction script, and time-corrected the arrival times to the Solar System barycenter using the position of \source{} \citep{2003ApJ...591L.143G}.

The first observation (00706396000) was taken in Photon Counting mode. 
For this observation, we used an annular source region with an inner radius of 4 pixels (7\arcsec) and an outer radius of 20 pixels (47\arcsec) to mitigate pile up\footnote{See \url{www.swift.ac.uk/analysis/xrt/pileup.php}}.
We extracted background events from an annulus of inner radius 64 pixels (150\arcsec) and outer radius 150 pixels (350\arcsec) centered on the source.

All other \swift\ observations were taken in Windowed Timing mode.
For these observations, to investigate the flux and spectral evolution of \source{}, we extracted a 20-pixel (47\arcsec) strip centered on the source. We extracted background events from a 50-pixel long (115\arcsec) 100 pixels away from the source.

We extracted the \swift-XRT spectra from the selected regions using {\tt extractor}, and fit using {\tt XSPEC} version 12.8.2\footnote{\url{http://xspec.gfsc.nasa.gov}}. Photons were grouped to ensure at least one photon was in each spectral bin. A summary of the {\it Swift} observations is found in Table~\ref{tab:obs}.

\subsection{{\it XMM}-Newton Observations}
\label{xmm}

We observed \src\ on 4 epochs with \xmm\ using the EPIC/pn and EPIC/MOS cameras in the Small Window mode, with time resolutions of 5.7\,ms and 0.3\,s, respectively. The epochs of {\it XMM} observations are listed in Table~\ref{tab:obs}. Here we use only the EPIC/pn data since they have better time resolution and sensitivity. We used the {\it XMM} Science Analysis System (SAS) version 16.0 and {\tt HEASOFT v6.19} to reduce the data. We pre-processed the raw Observation Data Files (ODF) using the SAS tool {\tt epproc} and filtered the event files so that single--quadruple events with energies between 0.1--12\,keV were retained, and standard ``FLAG'' filtering was applied. We extracted source events from an 18\arcsec\ radius region centered on \src{}. Background events were extracted from a 72\arcsec\ radius circular region placed away from the source.

\subsection{{\it NuSTAR} Observations}
\label{nustar}

We observed \src\ with \nustar\ on 5 epochs. We reduced the \nustar\ data with the {\tt nupipeline} scripts and {\tt HEASOFT v6.20}. We corrected the arrival times to the Solar System barycenter. Source events were extracted within a 1\arcmin\ radius around the centroid. Background regions were selected from the same detector as the source location, and spectra were extracted using the {\tt nuproducts} script. Using {\tt grppha}, channels 0--35 ($<3$ keV) and 1935--4095 ($> 79$ keV) were ignored, and all good channels were binned to have a minimum of one count per energy bin.

\begin{deluxetable}{lccc}
  \centering
  \tablecolumns{4} 
  \tablecaption{X-ray Observations of \psr\ in 2016. \label{tab:obs}}
  \tablewidth{0pt}
  \tabletypesize{\footnotesize}
  \tablehead{
    \colhead{Obs ID/Rev}   &
    \multicolumn{2}{c}{Obs Date} & 
    \colhead{Exp}\\
    \colhead{}  &
    \colhead{}                    &
    \colhead{}                    &
    \colhead{(ks)}                 
  }
  \startdata
  \sidehead{\swift-XRT}\hline
  00706396001      & 2016-07-28 --   & & \\
  00034632001--67  &   -- 2017-02-21 & & 159$^1$\\
  \hline
  \sidehead{\xmm}	\hline
  0741732601 & 2016-08-06  & & 21.6 \\
  0741732701 & 2016-08-15  & & 29.5 \\
  0741732801 & 2016-08-30  & & 34.0 \\
  0762032801 & 2016-12-13  & & 49.0 \\
  \sidehead{\nustar}  \hline         
  80102048002 & 2016-07-28  & & 54.4 \\
  80102048004 & 2016-08-05  & & 87.2 \\
  80102048006 & 2016-08-14  & & 95.4 \\
  80102048008 & 2016-08-30  & & 92.1 \\
  80102048010 & 2016-12-12  & & 94.3 \\
    \enddata
\tablenotetext{1}{Sum of all \swift\ exposure time used in this work.}
\end{deluxetable}

\section{Flux \& Spectral Evolution}

\subsection{Burst Search}
\label{sec:bursts}

\begin{figure}
	\includegraphics[width=\columnwidth]{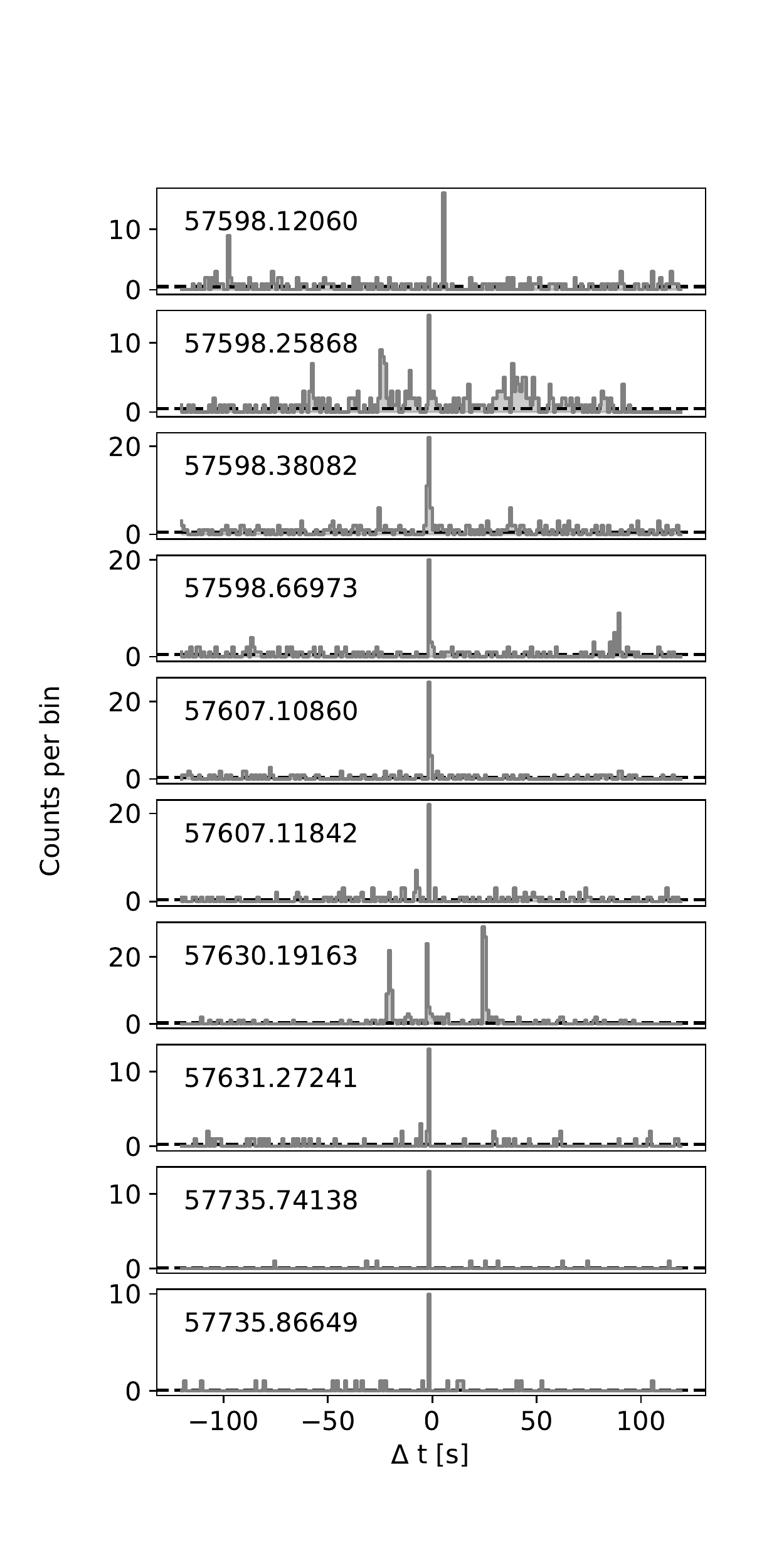}
	\caption{Bursts from \src{}.  Each panel shows a 3--79\,keV light curve binned at 1-s from the \nustar\ observations that contains burst emission. The average source count rate from each observation is plotted as a dashed black line. The count rates are not corrected for \nustar\ dead time. }
	\label{fig:bursts}
\end{figure}

We searched all X-ray observations of \newline \src{} presented here for magnetar-like bursts at timescales of 0.1\,s and 1\,s using the method presented in \cite{2011ApJ...739...94S} where  each Good Time Interval (GTI) is searched for statistically significant deviations from the mean count rate, assuming Poisson statistics.

In Figure~\ref{fig:bursts} we show the 3--79\,keV light curves surrounding each detected burst with a false alarm probability $P_{FA} \leq 10^{-6}$, with the time of the burst superimposed on the figure. 
Bursts that occurred in the same GTI are plotted together.
We note that the burst occurrence rate is highly clustered (i.e. non-Poissionian) -- of the 10 GTIs  in the \nustar\ data in which bursts were detected, only four had isolated, single bursts, whereas the other six contain multiple bursts. 
As every significant burst detected in \xmm\ was also detected with \nustar\, we do not present the \xmm\ light curves.

The shortest burst, on MJD 57335.86649 had a $T_{90}= 26 _{- 6 }^{+ 5 }$\,ms and contained 10 photons between 3--79\,keV.
This flux approached  the  $\sim$400 counts s$^{-1}$ maximum count rate that the {\it NuSTAR} detectors can process \citep{2013ApJ...770..103H}, and should therefore be taken as a lower limit on the fluences of the bursts.
We note that some of the brightest of these bursts, notably those on MJD 57630, were spectrally analyzed by \cite{2017ApJ...849L..20A}.
The remainder have insufficient counts for meaningful spectral analysis.
Regardless, all were removed from the data prior to the subsequent analysis.

\subsection{Long-term Flux and Spectral Evolution}

In this work, pulse-phase-averaged X-ray spectra were fit using \texttt{XSPEC} v12.9.0 \citep{1996ASPC..101...17A} with a common value for hydrogen column density ($N_\mathrm{H}$) for which we use  using \texttt{wilm} abundances \citep{2000ApJ...542..914W} and \texttt{vern} photoelectric cross-sections \citep{1996ApJ...465..487V}. We used Cash statistics \citep{1979ApJ...228..939C} for fitting and parameter estimation of the unbinned data. For the long-term evolution of the flux observed with the \swift{}-XRT, photon counting statistics limit us to fitting a single absorbed blackbody,  and $N_H$ has been fixed to $1.2\times10^{22}\,\mathrm{cm^{-2}}$ \citep{2016ApJ...829L..21A}.

We paired the \nustar\ and \xmm\ data that were gathered within a day of each other (see Table~\ref{tab:obs}) and simultaneously fit them with an absorbed blackbody plus power-law model. The bursts noted in \S\ref{sec:bursts} were excised from the event files. 

\begin{figure}
	\includegraphics[width=\columnwidth]{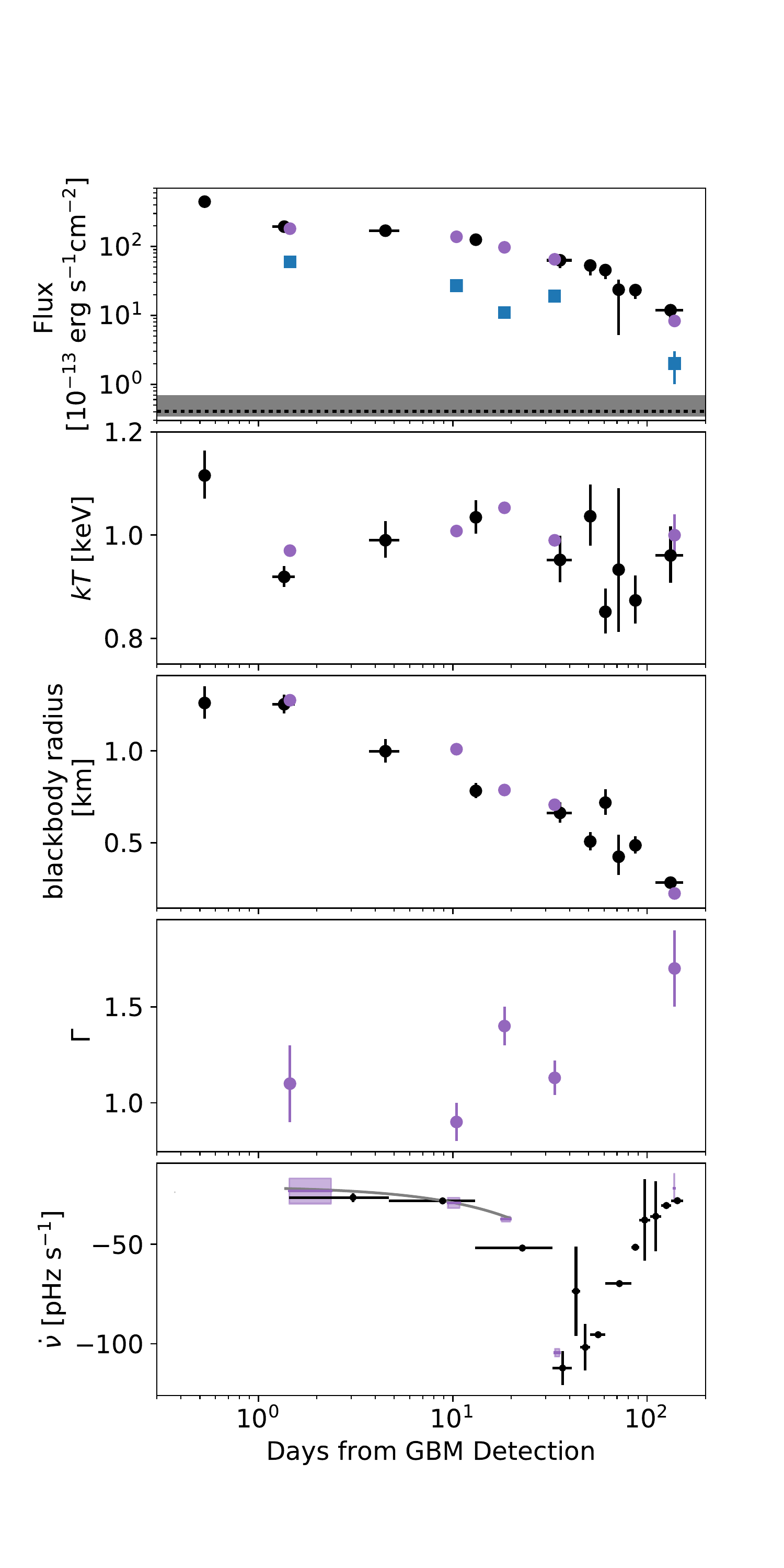}
	\caption{Spectral evolution of \src{}. The top panel shows the 0.5--10\,keV absorbed X-ray flux (circles), and the 10--79\,keV flux (blue squares). {\it Swift} data is in black, and the joint {\it XMM} and {\it NuSTAR} data are purple. The grey band in the quiescent 0.5--10\,keV flux, from \cite{2008ApJ...684..532S}.   The second panel shows the blackbody temperature. The third panel the implied blackbody radius assuming a distance of 8.4\,kpc \citep{2004MNRAS.352.1405C}. The fourth panel shows the hard power-law index. The bottom panel shows $\dot{\nu}$ (see \S\ref{sec:timing}).}
	\label{fig:swiftspec}
\end{figure}

\begin{figure*}
	\includegraphics[clip=True, trim=0.17in 0.13in 0.05in 0.02in, width=\textwidth]{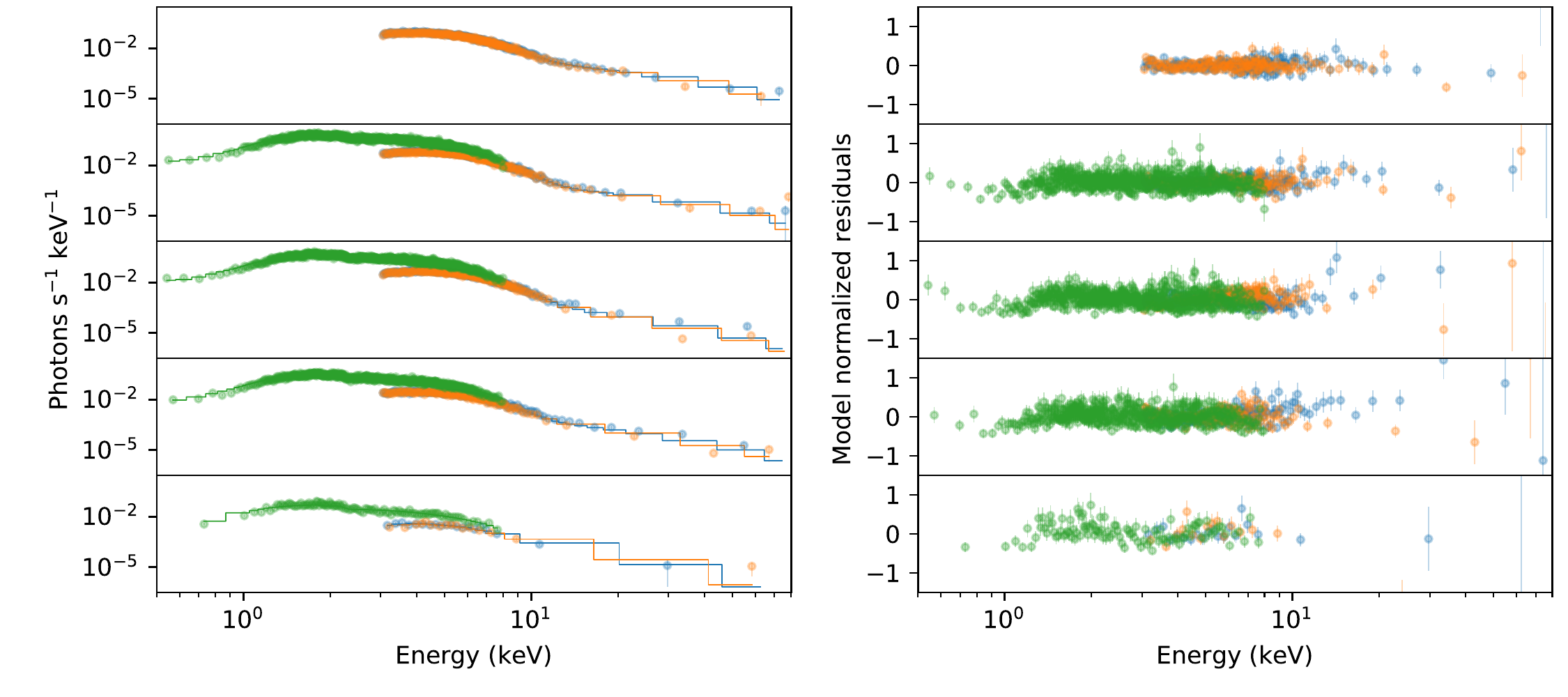}
	\caption{\nustar\ (orange and blue) and \xmm\ (green) spectra for \src{}. The left panels show the five epochs of \nustar\ observations starting from the earliest at the top. The right panels show the model normalized residuals after fitting the data with an absorbed blackbody plus power-law model (see Table~\ref{tab:swflux}.}
	\label{fig:nustarspec}
\end{figure*}

In Figure~\ref{fig:swiftspec} and Table~\ref{tab:swflux}, we present the long-term evolution of the 0.5--10\,keV and 10--79\,keV absorbed  flux, the blackbody temperature, the implied blackbody radius for a distance of 8.4\,kpc \citep{2004MNRAS.352.1405C}, as well as the hard power-law index. We co-fit closely spaced observations, grouping observations where the spectral parameters were consistent.
As is apparent in Figure~\ref{fig:swiftspec}, for the $\sim$20 days following the initial outburst observation, $kT$ rose modestly while the X-ray flux fell.
While this increase in $kT$ is modest, we detect it independently in both the \swift\ data, and the joint \xmm\ and \nustar\ data.
Figure~\ref{fig:nustarspec} shows the corresponding spectral fits to the \xmm\ and \nustar\ data.

\begin{deluxetable*}{lcccccc}
  \centering
  \tablecolumns{7} 
  \tablecaption{\swift-XRT, \xmm, and \nustar\ Spectral Results for \src{}. \label{tab:swflux}}
  \tablewidth{0pt}
  \tabletypesize{\footnotesize}
  \tablehead{
    \colhead{Date Range} &
    \colhead{$kT_\mathrm{BB}$} &
    \colhead{Radius\tablenotemark{a}} &
    \colhead{$F_\mathrm{BB}$\tablenotemark{b}} &
    \colhead{$\Gamma_\mathrm{PL}$}  &
    \colhead{$F_\mathrm{PL}$\tablenotemark{b}} &
    \colhead{C-stat/dof} \\
    \colhead{} &
 	\colhead{(keV)} & 
    \colhead{(km)} &
    \colhead{$(10^{-12}\,\mathrm{erg\,cm^{-2}\,s^{-1}})$} &
    \colhead{} &
    \colhead{$(10^{-12}\,\mathrm{erg\,cm^{-2}\,s^{-1}})$} & 
    \colhead{}
  }
  \startdata
  \sidehead{\swift-XRT} \hline
2016-07-27 &    1.12 $_{- 0.04 }^{+ 0.05 }$ &    1.26 $_{- 0.09 }^{+ 0.09 }$ & 45. $_{- 7.}^{+ 5. }$ & ... & ... & 320.0/388\\
2016-07-28 &    0.92 $_{- 0.02 }^{+ 0.02 }$ &    1.25 $_{- 0.05 }^{+ 0.05 }$ & 19.4 $_{- 1.8 }^{+ 1.6 }$ & ... & ... & 561.9/558\\
2016-07-31 -- 2016-08-02  &    0.99 $_{- 0.03 }^{+ 0.04 }$ &    1.00 $_{- 0.06 }^{+ 0.06 }$ & 17 $_{- 3 }^{+ 2 }$ & ... & ... &436.7/480 \\
2016-08-09 -- 2016-08-10 &    1.03 $_{- 0.03 }^{+ 0.03 }$ &    0.78 $_{- 0.04 }^{+ 0.04 }$ & 12.5 $_{- 2 }^{+ 1.6 }$ & ... & ... & 629.8/730\\
2016-08-26 -- 2016-09-06  &    0.95 $_{- 0.04 }^{+ 0.05 }$ &    0.66 $_{- 0.05 }^{+ 0.06 }$ & 6.3 $_{- 1.5 }^{+ 1.2 }$ & ... & ... & 442.6/509\\
2016-09-16 -- 2016-09-17  &    1.04 $_{- 0.06 }^{+ 0.06 }$ &    0.51 $_{- 0.05 }^{+ 0.05 }$ & 5.3 $_{- 1.5 }^{+ 0.9 }$ & ... & ... & 387.3/449\\
2016-09-26 -- 2016-09-27  &    0.85 $_{- 0.04 }^{+ 0.05 }$ &    0.72 $_{- 0.07 }^{+ 0.07 }$ & 4.5 $_{- 1.2 }^{+ 0.9 }$ & ... & ... & 305.6/383\\
2016-10-06 - 2016-10-07  &    0.93 $_{- 0.12 }^{+ 0.16 }$ &    0.42 $_{- 0.10 }^{+ 0.12 }$ & 2.4 $_{- 1.8 }^{+ 0.9 }$ & ... & ... & 145.2/164\\
2016-10-18 -- 2016-10-27  &    0.87 $_{- 0.04 }^{+ 0.05 }$ &    0.49 $_{- 0.05 }^{+ 0.05 }$ & 2.3 $_{- 0.6 }^{+ 0.4 }$ & ... & ... & 545.6/611\\
2016-11-15 -- 2016-12-28  &    0.96 $_{- 0.05 }^{+ 0.06 }$ &    0.28 $_{- 0.03 }^{+ 0.03 }$ & 1.2 $_{- 0.4 }^{+ 0.3 }$ & ... & ... & 1702.5/1800\\\hline
  \sidehead{\xmm\ and \nustar} \hline
  2016-07-28\tablenotemark{c}  & $0.970_{-0.010}^{+0.009}$ &   $1.28_{-0.03}^{+0.03}$ & $18.1_{-0.3}^{+0.3}$ & $1.1_{-0.2}^{+0.2} $ & $6_{-1}^{+1}$ & 1699.3/1901 \\ 
  2016-08-05 & $1.008_{-0.006}^{+0.006} $ &   $1.01_{-0.01}^{+0.01}$ & $13.8_{-0.1}^{+0.1}$ & $0.9_{-0.1}^{+0.1}$ & $2.7_{-0.3}^{+0.4}$ & 3540.8/3598 \\ 
  2016-08-14 & $1.053_{-0.007}^{+0.007} $ &   $0.79_{-0.01}^{+0.01}$ & $9.7_{-0.1}^{+0.1}$& $1.4_{-0.1}^{+0.1}$ & $1.1_{-0.2}^{+0.2}$ & 3603.4/3635 \\ 
  2016-08-30 & $0.990_{-0.009}^{+0.009}$ &    $0.71_{-0.01}^{+0.01}$ & $6.5_{-0.1}^{+0.1}$ & $1.13_{-0.08}^{+0.09}$ & $1.9_{-0.2}^{+0.3}$ & 3626.5/3639 \\ 
  2016-12-12 & $1.00_{-0.04}^{+0.04}$    &    $0.22_{-0.02}^{+0.02}$ & $0.83_{-0.06}^{+0.06}$ & $1.7_{-0.1}^{+0.2}$ & $0.2_{-0.1}^{+0.1}$ & 2412.33/2584 
  \enddata
  \tablecomments{The uncertainties specified are for a 1.6-$\sigma$ (90\%) confidence interval.}
  \tablenotetext{a}{Blackbody radius assuming a distance of 8.4\,kpc.} 
  \tablenotetext{b}{Absorbed flux from 0.5--10\,keV for the blackbody and from 10--79\,keV for the power law, respectively.}
  \tablenotetext{c}{\nustar-only fit.}
\end{deluxetable*}

\subsection{Pulse Profile}
\label{sec:profile}
\begin{figure*}
	\includegraphics[width=\textwidth]{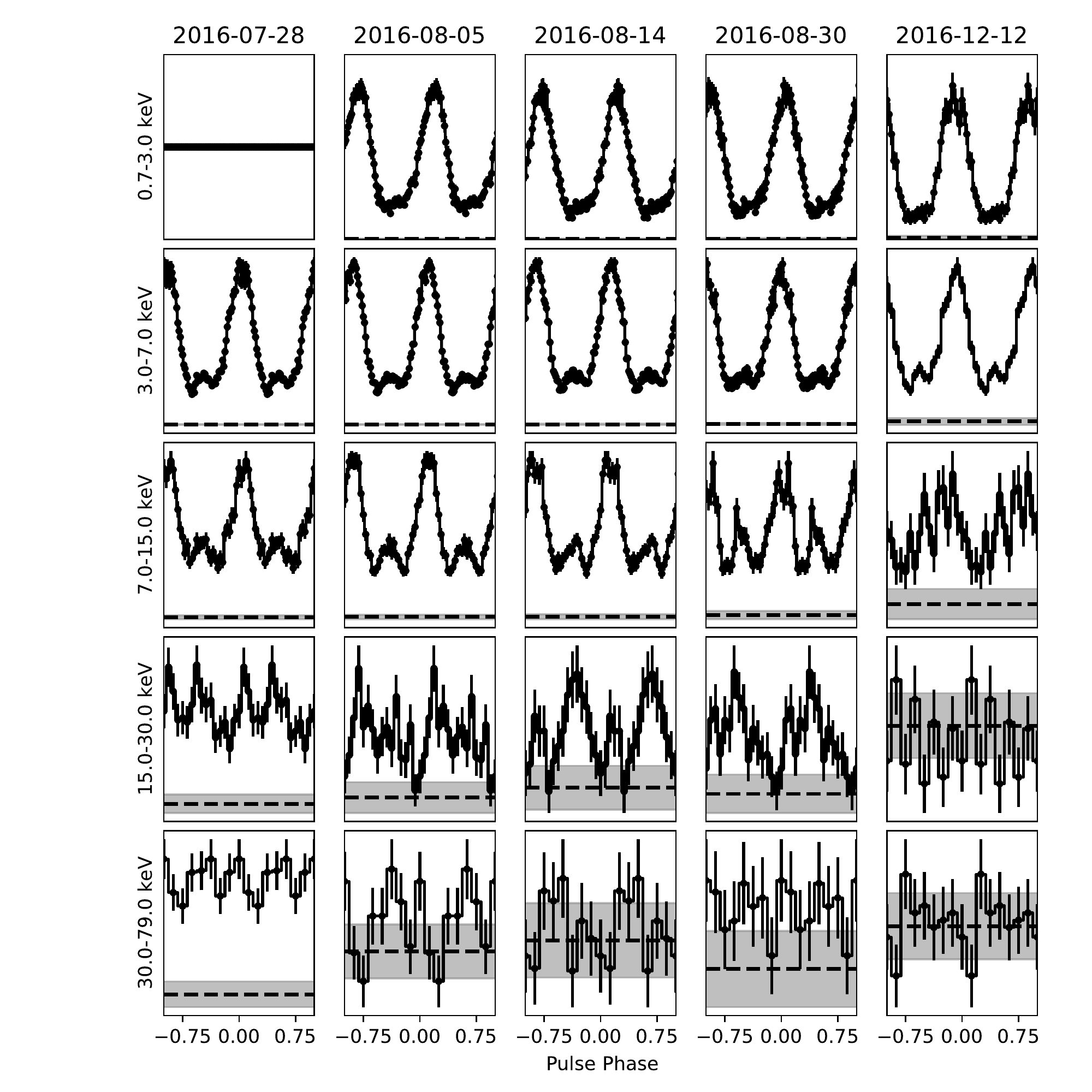}
	\caption{\xmm\ and \nustar\ pulse profiles of \src{}. The  panels show the five epochs of  observations starting from the earliest at the left, with energy increasing from the top panel to the bottom. The dashed black line shows the background count-rate with the uncertainty indicated by the grey band. The 0.7--3\,keV profiles are from \xmm\ and all other profiles are from {\it NuSTAR}. }
	\label{fig:pulse}
\end{figure*}

Using the timing solution derived as described in \S\ref{sec:timing}, we created pulse profiles from the \xmm\ and \nustar\ observations.
We removed all photons from times within 200\,s of a burst to avoid contamination of the pulse profile. 
These profiles are presented in Figure~\ref{fig:pulse}.
We calculated the root-mean-squared pulsed fraction of \src{} in several energy bands, using the method described by \cite{2015ApJ...807...93A}.
These pulsed fractions are presented in Table~\ref{tab:pulse}. 
Upper limits are given at the 99.9\% level, and the entry is blank if the upper limit is greater than 100\%.

In the soft X-ray band (below 3\,keV), the pulse shape and fraction are remarkably similar in all the observations, despite the roughly order of magnitude drop in the absorbed flux with time.  E.g. the pulse fraction only varies from $\sim67-75$\%; see Table~\ref{tab:pulse} for details.  These are comparable to the unusually high quiescent pulsed fraction of 74\% \citep{2005ApJ...630..489G} 

In the harder X-rays, the pulse shape is much more varied.
For the first four epochs, pulsed emission is detected up to 30\,keV.
Above 3\,keV, the profile develops a second peak (see Fig.~\ref{fig:pulse}) which persists until the fourth \nustar\ observations.

The pulse fraction in the 15--30-keV band increased following the outburst. 
Beginning with a pulse fraction of $34\pm6$\% on 2016-07-28, in all further \nustar\ observations, the pulse fraction is consistent with 100\% pulsed, and remained so until the source was too faint to detect.

\begin{table*}
\begin{center}
    \caption{Pulse Fractions of \src{}\tablenotemark{$a$}\tablenotemark{$b$}}
    \label{tab:pulse}
\begin{tabular}{cccccc}

	  & 2016-07-28 & 2016-08-05 & 2016-08-14 & 2016-08-30 & 2016-12-12 \\ 
    \hline 
	0.7-3\,keV & ... & 67$\pm$1 & 71$\pm$1 & 69$\pm$1 & 75$\pm$3 \\ 
	3-7\,keV & 62.1$\pm$0.7 & 59.7$\pm$0.6 & 57$\pm$1 & 55$\pm$1 & 62$\pm$3 \\ 
	7-15\,keV & 44$\pm$2 & 48$\pm$2 & 50$\pm$2 & 44$\pm$2 & 56$\pm$14 \\ 
	15-30\,keV & 34$\pm$8 & 80$\pm20$ & 100${_{-20}}$ & 95$^{+5}_{-20}$ & ... \\ 
	30-79\,keV & $<$40 & ... & ... & ... & ... \\ 
	\hline 
	\hline 
    
\end{tabular}
\end{center}
\tablenotetext{a}{Pulsed fractions expressed as percentages.}
\tablenotetext{b}{Entries left blank are unconstrained.}
\end{table*}

\section{Timing Analysis}
\label{sec:timing}

\begin{figure}[p]
	\includegraphics[width=\columnwidth]{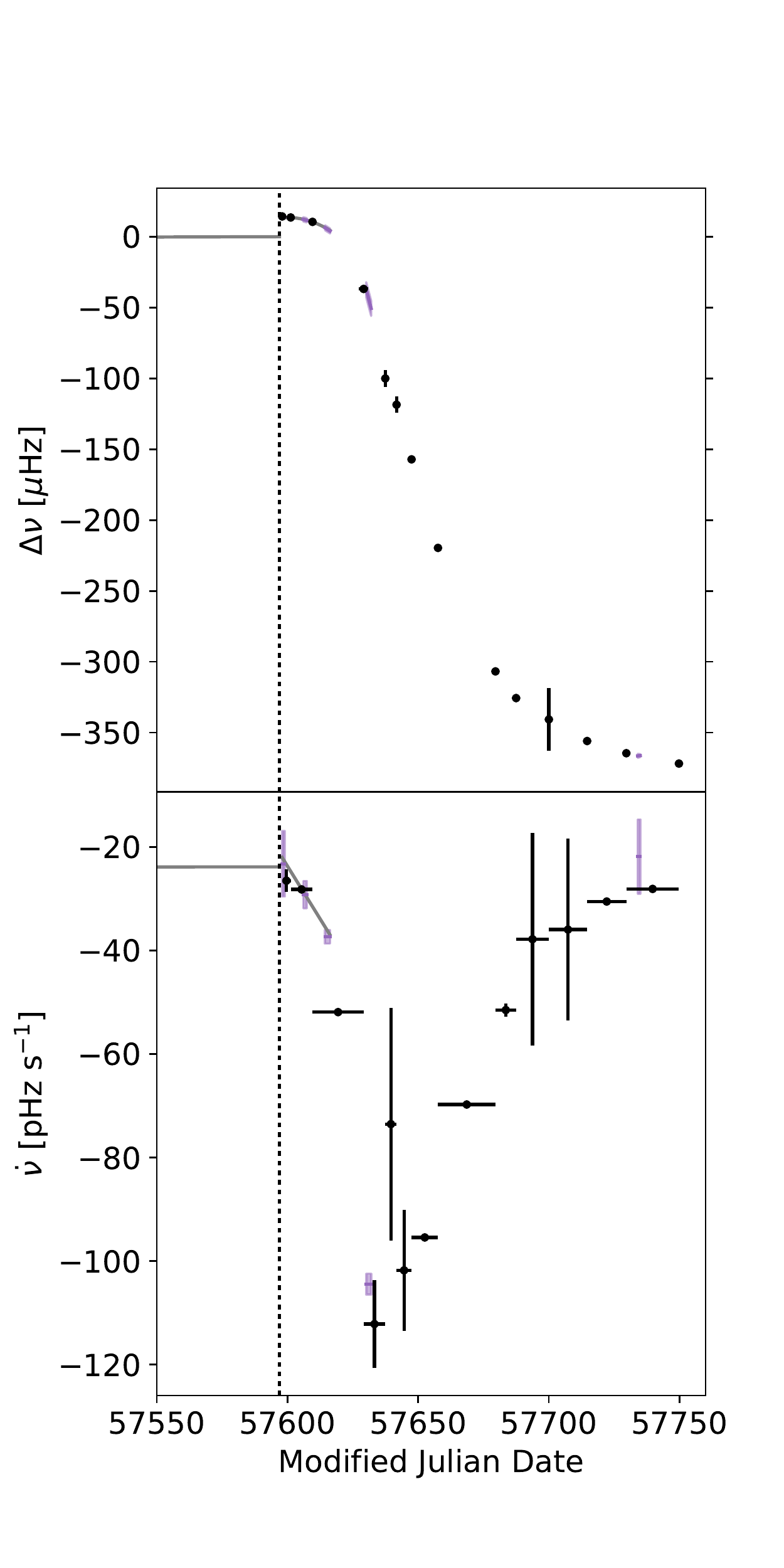}
	\caption{Spin evolution of \source{} near the outburst.  The top panel shows the spin frequency ($\nu$) evolution with the pre-outburst ephemeris subtracted. The bottom panel shows the spin-down rate ($\dot{\nu}$) over this same period. The solid grey lines indicate phase-coherent timing solutions, with the pre-outburst solution coming from \cite{2016ApJ...829L..21A}. Short-term phase connected solutions from the \nustar{} and \xmm{} observations are depicted in purple. Finally, the black data points in the top panel are from \swift{} using the maximum likelihood method; see \S\ref{sec:timing}. The black points in the bottom panel show the average spin-down rate ${\dot{\nu}}$ between any two \swift{} measurements.  The error bars are plotted, and when not seen, are comparable to the size of the point.}
	\label{fig:timing}
\end{figure}

\begin{table}
    \begin{center}
    \caption{Phase-Coherent Ephemeris for \src{} following the outburst.}
    \label{tab:timing}
    \begin{tabular}{ll}
    \hline
    \hline
    \multicolumn{2}{c}{Post-Outburst Ephemeris} \\
    Dates (MJD)         & 57597.72--57616.3\\
    Dates               & 28 July --16 Aug 2016 \\
    Epoch (MJD)         & 57600.\\
    $\nu\;$ (s$^{-1}$)            &  2.439 837 24(8)\\
    $\dot{\nu}\;$ (s$^{-2}$)            &$-$2.36(1)$\times 10^{-11}$\\
    $\ddot{\nu}\;$ (s$^{-3}$)            &$-$9.7(2)$\times 10^{-18}$\\
    RMS residual (ms) & 6.2\\
    RMS residual (phase) & 0.015\\
    $\chi^2_\nu$/dof & 1.98/168 \\
    \hline
    \hline
    \end{tabular}
    \newline

    Note: Figures in parentheses are the  1$\sigma$ \textsc{tempo2} uncertainties in the least-significant digits quoted. The source location was fixed at the {\it Chandra} position for the timing analysis  \citep{2003ApJ...591L.143G}. 
    \end{center}
\end{table} 

In addition to the spectral work described above, we conducted a timing analysis of \src{}.
For the first 19-days of the outburst, during which the source's X-ray flux was greatly enhanced and well sampled, we employed a phase-coherent analysis.
To do this, we extracted pulse times-of-arrival (TOAs) using the maximum likelihood timing method described by \cite{2009LivingstoneTiming}.
To optimize the signal-to-noise ratio of the pulse profiles, photons above 1.1\,keV were used for \swift{} and \xmm{} and photons from 3--25\,keV for \nustar{}.
These TOAs were then fit to a standard pulsar timing model using the {\tt tempo2} pulsar timing package \citep{2006MNRAS.369..655H}.
 
In Table~\ref{tab:timing}, we present a phase-coherent solution valid in the interval MJD 57597--57616. 
Note the high $\chi^2$ value, and indicator of  a high amount of timing noise even within this 19-day period. 
Over these 19 days, the spin-down rate increased from $-2.61(5)\times 10^{-11}$\,Hz\,s$^{-1}$ to $-3.8(1)\times 10^{-11}$\,Hz\,s$^{-1}$.

Due to this rapid evolution of $\dot{\nu}$, we were unable to maintain a phase-coherent timing solution past this initial period of the post-glitch timing evolution.

We then employed the maximum likelihood method of period-finding described by \cite{2018ApJ...852..123F} wherein the standard maximum likelihood timing method from \cite{2009LivingstoneTiming} is extended to both a trial reference phase $\delta$ and trial rotation frequency $\nu$.
The resulting two-dimensional probability density is then given by 
\begin{equation}
	\label{eqn:2d_pdf}
	\mathrm{Prob}(\nu,\delta) = \prod_{i=1}^N I\left(\phi_i(\nu) - \delta\right)\,,
\end{equation}
where $I$ is the probability density function created from a high signal-to-noise template profile.
This product is calculated over a finely sampled\footnote{Our search grid went from +100~$\mu$Hz to -1000~$\mu$Hz from the pre-outburst solution, motivated by the frequency measurements from \nustar{} and \xmm{}.}, frequency grid, and marginalized over $\delta$ to obtain a measurement of the true spin frequency and corresponding uncertainty.

For the joint \nustar{} and \xmm{} observations, the observation duration and the high signal-to-noise allowed the creation of short-term phase coherent measurements of both $\nu$ and $\dot{\nu}$. 
As the \nustar{} and \xmm{} were observed close in time to each other, TOAs extracted from each telescope were combined to create timing solutions.
These timing solutions are shown in Figure~\ref{fig:timing} as purple lines with a corresponding region showing the 68\% uncertainty region.

In Figure~\ref{fig:timing}, we summarize the spin-frequency evolution of \src{} around the epoch of the  outburst.  
The evolution can be described by a $(1.40\pm2)\times10^{-5}$\,Hz spin-up glitch at the time of the outburst\citep{2016ApJ...829L..21A}, followed by a period of increased and highly variable spin down, resulting in a net spin-down of the pulsar of $\sim 3.5\times10^{-4}$\,Hz by the end of the observing campaign. This corresponds to a very large fractional frequency change of 
$\Delta\nu/\nu \simeq -1.4 \times 10^{-4}$, or an over-recovery of the initial spin-up glitch by a factor of 24.  This represents
an equivalent frequency change to that expected in $\sim$6 months of standard spin-down at the nominal $\dot{\nu}$.

\section{Discussion}

We have reported on the observational aftermath of the 2016 magnetar-like outburst from the young, highly magnetized radio pulsar \src{}.
We have shown that the time scales for the evolution of the soft and hard X-ray components are similar.
We detect a brief $\sim$10-day interval during which the flux was falling, but the spectrum was hardening, in contrast
to the standard flux/hardness correlation usually seen in magnetar outburst relaxations.
Moreover we have shown that the outburst hard X-ray emission was
highly pulsed and included a second, heretofore unseen hard component above 3 keV.
Further, we examined the relaxation following a spin-up glitch and showed
that this pulsar experienced a period of intense
spin-down that greatly overcompensated for the initial spin-up event, eventually resulting in a massive net spin down of magnitude
$\Delta\nu \simeq -3.5 \times 10^{-4}$~Hz ($\Delta\nu/\nu \simeq -1.4 \times 10^{-4}$), in addition to erratic spin-down variations throughout.

\subsection{Energetics}

The quiescent X-ray output has been observed to
have luminosity of $0.9 \times 10^{33}$~erg~s$^{-1}$ in
the 0.5--10~keV band \citep{2005ApJ...630..489G}, assuming a distance
of 8.4~kpc \citep{2004MNRAS.352.1405C}, only 0.0009 of the
pulsar's spin-down luminosity, $\dot{E}=2.3 \times 10^{36}$~erg~s$^{-1}$.  Thus the quiescent emission
can be fully accounted for in the spin-down budget.
Nevertheless, the somewhat high $kT$ of this emission, together with its high pulse fraction,
prompted \citet{2005ApJ...630..489G} to suggest some form of active interior magnetic heating.  It is interesting to ask whether
the outburst emission luminosity exceeds at any time
$\dot{E}$, which would constitute additional evidence
for an interior energy source in addition to the rotational
kinetic energy.

On 2016 July 27, the day of the observation with the highest flux (see Table~\ref{tab:swflux}), the total 0.5-10-keV flux was 3.8$\times 10^{35}$~erg~s$^{-1}$, corresponding to 0.16$\dot{E}$. 
By the next day, when the first \nustar\ observation occured, the total 0.5-79-keV flux was 2.0$\times 10^{35}$~erg~s$^{-1}$, corresponding to 0.09$\dot{E}$, with the hard-band flux accounting for 1/3 of the soft-band flux.  
The hard component, however, had a very
flat spectrum with $\Gamma_{PL} = 1.1$ and no evidence for a cutoff.  Magnetars with hard spectral components in general 
have not had any cutoffs measured, with the exception of
4U 0142+61, for which a cutoff energy of $\sim$300~keV
has been observed \citep{2008A&A...489..245D}.  If the cutoff for \src{} were 300\,keV, the hard-band flux would be just 40\% higher.
For the total flux to correspond to a luminosity  equal to $\dot{E}$, the cutoff energy would have had
to have been $\sim$4\,MeV.  This may be testable one day with a future soft gamma-ray observatory such as ComPair \citep{2015arXiv150807349M} or AMEGO \citep{2017JInst..12C1024R}, if another outburst is observed.

Interestingly, the overall unabsorbed flux increase of a factor of $\sim$300 relative to quiescence represents one of the largest dynamic ranges yet seen in any magnetar outburst, \cite[see][]{2018MNRAS.474..961C}. 
It is interesting that the  highest increases in energy output are also from sources having the lowest spin-inferred magnetic field strengths, and among the lowest quiescent luminosities. 
This is especially interesting when considering that the sources with the highest inferred fields
(e.g. SGR\,1900+14 or 1E\,1841$-$045) have shown flux increases of well under a factor of 10.
This correlation was studied in detail
by \citet{2018MNRAS.474..961C} and has been interpreted \citep{2012ApJ...750L...6P}
as being a result of a natural limit to the X-ray luminosity of any magnetar, due to neutrino emissivity limiting the maximum temperature of the crust.  In this case, the maximum dynamic range should be determined by the quiescent flux \citep[see also][]{2016ApJ...833..261B}.  The large flux increase of
\src{} in outburst is further support of this proposed picture.

\subsection{Soft X-ray Flux Evolution}

The soft-X-ray flux evolution of \src{} is typical for magnetar outbursts \citep[e.g.][]{2011A&A...529A..19B, 2004ApJ...609L..67G, 2012ApJ...757...68A}.
This evolution is usually interpreted as either  passive cooling of the crust of the neutron star following the initial energy injection \citep[e.g.][]{2009ApJ...698.1020B,  2012ApJ...761...66S, 2015ApJ...809L..31D}, or by the gradual untwisting of large currents or ``j-bundles'' in the stellar magnetosphere which were formed in response to a shear deformation of the surface \citep[e.g][]{2009ApJ...703.1044B}.
As crustal cooling models have shown success in reproducing the flux evolution post-outburst
in magnetars \cite[e.g.][]{2018arXiv180400266A}, and that of \src{} is similar to others, likely crustal cooling is viable in this case as well, though such fitting is beyond the scope of this paper.  However the monotonically shrinking blackbody radius seems at odds with a crustal cooling model in which a initial, localized energy injection presumably spreads over the surface in time.  On the other hand, the shrinking blackbody radius is expected in j-bundle relaxation models, since this radius is dominated by the shrinking emission from the footpoints of the bundle \citep{2017ApJ...844..133C}

The possible anti-correlation between $kT$ (rising) and flux (falling) in the first $\sim$10 days
post-outburst (see Fig. 1) is the reverse
of the standard hardness/flux correlation generally seen in magnetar outbursts.  There
is evidence for similar behavior in at least
one other source \citep{2014ApJ...786...62S}.
Moreover, the similar time scale of erratic behavior in the evolution of $\dot{\nu}$ (see Fig.~\ref{fig:swiftspec}) is intriguing.  It is suggestive of a relationship between the torque as determined by field lines near the light cylinder and the blackbody emission, suggesting the latter does not completely emerge from the surface.
This warrants further investigation.

\subsection{Hard X-ray Flux and Evolution}

The \src{} outburst represents the first observation of the time evolution of a hard X-ray component
in a magnetar outburst.
We have observed that the hard X-ray tail abates on a time scale
comparable to that of soft X-ray emission: several months. This is in agreement with the expectations of the model predictions of \citet{2013ApJ...762...13B}, in which the two are closely coupled, being produced by j-bundle untwisting, albeit with the hard X-rays from the top of the bundle and the soft X-rays from the footpoints.  On the other hand, in the case of \src{}, the outburst emission has luminosity well below that available from spin-down, hence need not be powered by magnetic activity as is assumed by \citet{2013ApJ...762...13B}.
There may be more conventional mechanisms to explain these observations, as suggested by
\citet{2009A&A...501.1031K} in
consideration of a similar outburst from the high-B rotation-powered pulsar PSR J1846$-$0258 \citep{2008Sci...319.1802G}.

The emergence of a new hard pulsed component (see Fig.~\ref{fig:pulse}) is also interesting.
The high pulsed fractions seen for the hard X-ray emission (Table~\ref{tab:pulse}) are,
in addition to being similar to those for
other magnetars in the same energy
range \citep[e.g.][]{2006ApJ...645..556K},
also in agreement with the predictions of the j-bundle
relaxation model, though viewing angle and magnetic inclination angle play a role. These data
could thus be valuable for modeling similar to that done by
\citet{2014ApJ...786L...1H}, although this requires phase-resolved spectral modeling for which our limited energy range is likely a hindrance.  Still this seems promising given the well constrained geometry of this pulsar
from radio polarization and
$\gamma$-ray light-curve modeling
\citep{2003ApJ...590.1020C, 2011ApJ...743..170P}.

\subsection{Timing Behavior}

Glitches -- sudden increases in the spin frequency of a radio pulsar -- are particularly common in young pulsars \citep[e.g.][]{2011MNRAS.414.1679E}, including in magnetars \citep[e.g.][]{2008APJ.673..1044D, 2017ARA&A..55..261K}. For the most part, glitches in pulsars and magnetars are similar, being of comparable amplitude and having similar occurrence rates. However, magnetar glitches are often accompanied by radiative outbursts, unlike the vast majority of radio-pulsar glitches \citep[e.g.][]{2008APJ.673..1044D}. Moreover, magnetar glitches sometimes show unique timing behavior:  a typical spin-up glitch followed by a temporary increase in the spin-down rate of the pulsar, leading to a net spin-down of the pulsar \citep[see][and references therein]{2017ApJ...834..163A}.

\src{} itself has, in the past, exhibited atypical pulsar glitches.  In 2007, it had a glitch accompanied by a change in radio pulse profile \citep{2011MNRAS.411.1917W}, the first evidence for radiative changes in a rotation-powered pulsar at a glitch epoch.  
This glitch  may have also been accompanied by a change in braking index \citep{2015MNRAS.447.3924A}.

The initial stages of the glitch associated with the 2016 outburst of \src{}, first presented in \cite{2016ApJ...829L..21A}, were typical among radio pulsar glitches.
However, as reported here, the rapid increase in $\dot{\nu}$ and subsequent variation and decline in magnitude is not a behavior seen in glitches in radio pulsars.
That is, this recovery behavior is very poorly described by an exponential decay typical in radio pulsar glitch recoveries.

This rapidly changing $\dot{\nu}$ is similar to that observed following several magnetar outbursts -- for example in 1E\,1048.1$-$5937 \citep{2004ApJ...609L..67G, 2009ApJ...702..614D, 2015ApJ...800...33A}, in  PSR J1622$-$4950 \citep{2017ApJ...841..126S, 2018arXiv180401933C}, and in 1E\,1547.0$-$5408 \citep{2012ApJ...748....3D}.
In all these cases, it seems likeliest that the torque is exerted by variation of field line structure in the outer magnetosphere, near the light cylinder, as this should yield the largest lever-arm.
It is challenging to envision an internal origin for such torques as this would require a large fraction of the stellar moment of inertia coupling and decoupling on short time scales.

Thus, the similarity of the early stages of \src{}
glitch behavior in the 2016 outburst to that of typical radio pulsar glitches suggests a common origin {\it interior} to the star, likely with conventional vortex line unpinning and related transfer of superfluid angular momentum to the crust.  On the other hand, in high-B objects, this glitch couples with the external field,
likely via movement of crustal footpoints \citep[e.g.][]{2013ApJ...774...92P}, such that the relaxations in large magnetar and magnetar-like events are dominated by processes in the outer magnetosphere, unlike in lower-B sources.  The magnetar RXS\,J1708$-$4009 has exhibited radiatively quiet glitches, all of which have had more conventional, radio-pulsar like recoveries \citep[e.g.][]{2003ApJ...596L..71K, 2014ApJ...783...99S}.  This suggests modest internal glitches that did not disturb the crust sufficiently to result in magnetospheric anomalies that would produce either radiation or external relaxation torques.

The other high-B pulsar which exhibited magnetar-like behavior, PSR\,J1846$-$0228, also had a glitch at the time of the outburst which over-recovered by a factor of $\sim$9 \citep{2010ApJ...710.1710L}.  By comparison,
for \src{}, the over-recovery factor was 24.
This PSR\,J1846$-$0258 glitch was well fit by an exponential decay in $\dot{\nu}$  -- unlike  the more erratic behavior we report here.  However, we argue that the large over-recovery originated in the outer magnetosphere, where little to no memory need exist of the magnitude of the original internal spin-up glitch.  This suggests more generally that for high-B sources in which glitches are accompanied by large radiative outbursts, large over-recoveries and/or erratic timing variations should be generic.  Conversely, radiatively silent magnetar glitches in this picture should typically have simple timing relaxation, with no large over-recoveries.  Careful monitoring of future magnetar and high-B radio pulsar glitch and radiative behavior can test these ideas.

\section{Conclusions}
We have presented the observational aftermath of a magnetar outburst from the young, high-magnetic-field radio pulsar \src{}.
We have shown that many of the properties of the post-outburst relaxation are consistent with 
 models of magnetospheric twisting accompanied by crustal shearing due to internal magnetic-field related forces.  Among predicted behaviors we have confirmed are a similarity in the time scales of the decays of the soft and hard X-ray components, with both being $\sim$months.
Moreover we have shown that the outburst hard X-ray emission in \src{} was
highly pulsed, with pulsed fraction as high as $\sim$100\%, typical for magnetars, and expected in models of cooling of twisting currents in the outer magnetosphere.
We have also reported on erratic timing behavior following an initial spin-up glitch, resulting in a substantial over-recovery, 
 eventually resulting in a massive anti-glitch of magnitude
$\Delta\nu/\nu \simeq -1.4 \times 10^{-4}$ or
an over-recovery of over a factor of 20.
We argue that such behavior may be generic in magnetar and high-B pulsar glitches involving radiative outbursts, as for such events an internally generated spin-up may result in crustal shearing that communicates with the external magnetosphere, where large recovery torques having little to no memory of the original event can be supplied. 

\acknowledgements
This research has made use of data obtained through the High Energy Astrophysics Science Archive Research Center Online Service, provided by the NASA/Goddard Space Flight Center.
This work made use of data from the {\it NuSTAR} mission, a project led by the California Institute of Technology, managed by the Jet Propulsion Laboratory, and funded by the National Aeronautics and Space Administration.
This work made use of observations obtained with \xmm\, an ESA science mission with instruments and contributions directly funded by ESA Member States and NASA.
V.M.K. is supported by a Lorne Trottier Chair in Astrophysics \& Cosmology, a Canada Research Chair, CIFAR, by an NSERC Discovery Grant and Herzberg Award, and by FRQNT/CRAQ.
R.F.A. acknowledges support from an  NSERC  Postdoctoral Fellowship.
P.S. is a Covington Fellow at DRAO.

\facilities{\swift{}, \nustar{}, \xmm{}.}
	
\software{\newline {\tt astropy} \citep{2013A&A...558A..33A},\newline
				   {\tt xspec} (\url{http://xspec.gfsc.nasa.gov}),\newline
				   {\tt heasoft} (\url{https://heasarc.nasa.gov/lheasoft/})} 

\bibliographystyle{yahapj}
\bibliography{rfa,vmk}
\end{document}